\documentclass[twocolumn]{article}

\usepackage{hyperref}
\hypersetup{colorlinks=true,linkcolor=blue,urlcolor=blue,citecolor=blue}
\usepackage[a4paper,margin=20mm]{geometry}
\usepackage{amssymb}
\usepackage{amsmath}
\usepackage{graphicx}
\usepackage{flushend}
\usepackage{cite}
\usepackage{fancyhdr}
\title{\large \bf The origin of $g\approx4$ EPR line in magnetic nanocomposites: Manifestation~of~double quantum transitions in ferromagnetic granules}

\author{\normalsize \bf A.\,B. Drovosekov$^a$, M.\,Yu. Dmitrieva$^{a,b}$, A.\,V. Sitnikov$^{c,d}$, S.\,N. Nikolaev$^d$, V.\,V. Rylkov$^{d,e,f}$}

\date{\small \it $^a$Kapitza Institute for Physical Problems, Russian Academy of Sciences, Moscow, 119334 Russia \\ $^b$National Research University Higher School of Economics, Moscow, 101000 Russia \\ $^c$Voronezh State Technical University, Voronezh, 394026 Russia \\ $^d$National Research Centre Kurchatov Institute, Moscow, 123182 Russia \\ $^e$Institute of Theoretical and Applied Electrodynamics, Russian Academy of Sciences, Moscow, 125412 Russia \\ $^f$Kotelnikov Institute of Radio Engineering and Electronics, Fryazino Branch, Russian Academy of Sciences,\\ Fryazino, Moscow region, 141190 Russia}

\begin{document}

\twocolumn[
\begin{@twocolumnfalse}
\maketitle
\begin{abstract}
\noindent Films of metal-insulator nanogranular composites M$_x$D$_{100-x}$ with different compositions and atomic percentage of metal and dielectric phases (M = Fe, Co, Ni, CoFeB; D = Al$_2$O$_3$, SiO$_2$, ZrO$_2$; $x\approx15{-}60$~at.\,\%) are investigated by electron magnetic resonance in a wide range of frequencies ($f=7{-}37$~GHz) and temperatures ($T=4.2{-}360$~K). At concentrations of the metallic ferromagnetic phase below the percolation threshold, the experimental spectra, besides the conventional ferromagnetic resonance signal, demonstrate an additional absorption peak characterized by a double effective g-factor $g\approx4$. The appearance of such a peak in the resonance spectra and its unusual properties are explained in the framework of the quantum mechanical ``giant spin'' model by the excitation of ``forbidden'' (``double quantum'') transitions in magnetic nanogranules with a~change of the spin projection $\Delta m = \pm2$.
\vskip 6mm
\end{abstract}
\end{@twocolumnfalse}
]

\thispagestyle{fancy}

\noindent
\textbf{1. Introduction}
\vskip 2mm

Magnetic nanoparticles and nanogranular systems have long been a subject of intensive research, due to unusual physical properties of these objects, as well as the wide possibilities of applications \cite{Dormann1992, Gubin2009, Bedanta2013}. From the viewpoint of fundamental physics, magnetic nanoparticles can be considered as an intermediate case between paramagnetic (PM) ions and macroscopic ferromagnets. In particular, ensembles of such particles exhibit so-called ``superparamagnetic'' properties, while the magnetic dynamics of individual nanoparticles in some cases can be described within the framework of both classical and quantum approaches \cite{Noginova2007, Atsarkin2020}.

Magnetic metal-insulator nanogranular composites (nanocomposites) are an array of ferromagnetic (FM) nanogranules randomly distributed in a solid insulating medium (matrix). In previous works \cite{DrovJMMM, DrovJETPlett, Drov2022-1, Drov2022-2, Drov2023}, we studied nanocomposite films of different compositions M$_x$D$_{100-x}$ based on transition FM metals M~= Fe,~Co,~CoFeB and dielectrics D = Al$_2$O$_3$, SiO$_2$, LiNbO$_3$. The value $x$ in the formula M$_x$D$_{100-x}$ reflects the nominal atomic content of the metal phase in the nanocomposite, a significant part of which forms FM nanogranules. At the same time, some part of the FM phase proves to be dispersed in the form of PM ions Fe and Co in the insulating space between the granules \cite{Rylkov2019, Rylkov2020}.

In the works \cite{Drov2022-1,Drov2022-2,Drov2023} nanocomposite films M$_x$D$_{100-x}$ were studied by magnetic resonance in a wide range of frequencies ($f=7{-}37$~GHz) and temperatures ($T=4.2{-}360$~K). In addition to the usual ferromagnetic resonance (FMR) signal, the experimental spectra demonstrated an additional much weaker absorption peak characterized by an effective g-factor $g\approx4.3$. Note that a similar signal is often observed in studies of iron-based nanoparticles in various non-magnetic media \cite{Koksharov2001, Jitianu2002, Edelman2012} and is associated with the electron paramagnetic resonance (EPR) of isolated Fe$^{3+}$ ions present in the system \cite{Castner1960, Wickman1965, Kliava1988}. In some cases, the $g\approx4.3$ EPR peak is also manifested for Co$^{2+}$ ions \cite{Abragam1972, Legein1993, Raita2011}. However, unlike the traditional EPR of Fe$^{3+}$ and Co$^{2+}$ ions, in our case the observed peak demonstrates a number of unusual properties:

\noindent
1) It is much better manifested in the longitudinal geometry of the resonance excitation;

\noindent
2) With a change of the FM phase concentration $x$, the temperature dependence of the peak intensity $I(T)$ is modified: at low $x$, this dependence has a decreasing character, but when approaching the percolation threshold it becomes increasing.

Note that these unusual features of the additional peak do not allow to associate it with inhomogeneities of the samples, or with the excitation of inhomogeneous magnetic modes in the films, as, for example, in the works \cite{Wang1995, Lesnik2003, Gomez2004, Kakazei2005, Pires2006, Vyzulin2006, Kakazei2015, Martyanov2015, Kablov2016, Neugebauer2020, Kotov2020, Denisova2018}.

The present work is devoted to further investigation of the nature of the anomalous resonance peak. It is found that, in addition to nanocomposites based on FM metals Fe and Co, the $g\approx4.3$ EPR peak is also manifested for systems Ni$_x$D$_{100-x}$ (D = Al$_2$O$_3$, ZrO$_2$) based on pure nickel. Thus, the appearance of this peak in the resonance spectra cannot be explained by the presence of isolated Fe$^{3+}$ or Co$^{2+}$ ions in the system. In~view of this, an alternative explanation of the observed effects is proposed, suggesting a quantum mechanical approach to describing the magnetic resonance spectra of FM nanogranules \cite{Noginova2007, Atsarkin2020}.

\vskip 5mm
\noindent
\textbf{2. Samples and experimental method}
\vskip 2mm

Nanocomposite films M$_x$D$_{100-x}$ with a thickness of about $1{-}3~\mu$m were synthesized by ion beam sputtering on glass-ceramic substrates using composite targets \cite{Stognei2014, Kalinin2007}. The target represents a plate of FM metal Fe, Co, Ni or an alloy Co$_{40}$Fe$_{40}$B$_{20}$ (CoFeB), on which a set of rectangular strips of oxides Al$_2$O$_3$, SiO$_2$ or ZrO$_2$ are placed. The uneven arrangement of dielectric strips on the target surface allows the formation of a nanocomposite film M$_x$D$_{100-x}$ with a smooth controlled change in the concentration $x$ along the substrate in a wide range $\Delta x\approx30{-}40$~at.\,\%. Further studies are carried out on small pieces of the grown film with a size of $5\times5$~mm$^2$, so that the change of $x$ within one sample is less than 1~at.\,\%. The content of the metal phase in the films was determined by energy dispersive X-ray microanalysis.

According to transmission electron microscopy and X-ray diffraction data, the obtained composites consist of crystalline FM nanogranules randomly distributed inside the amorphous oxide matrix \cite{Kalinin2007, Rylkov2017, Stognei2010, Tregubov2013, Stognei2016, Filatov2017}. The granules have approximately spherical shape and their average size (2--8~nm) increases smoothly with a growth of the FM phase content in the nanocomposite (Fig.~1).

\begin{figure}
\centering
\includegraphics[width=0.68\columnwidth]{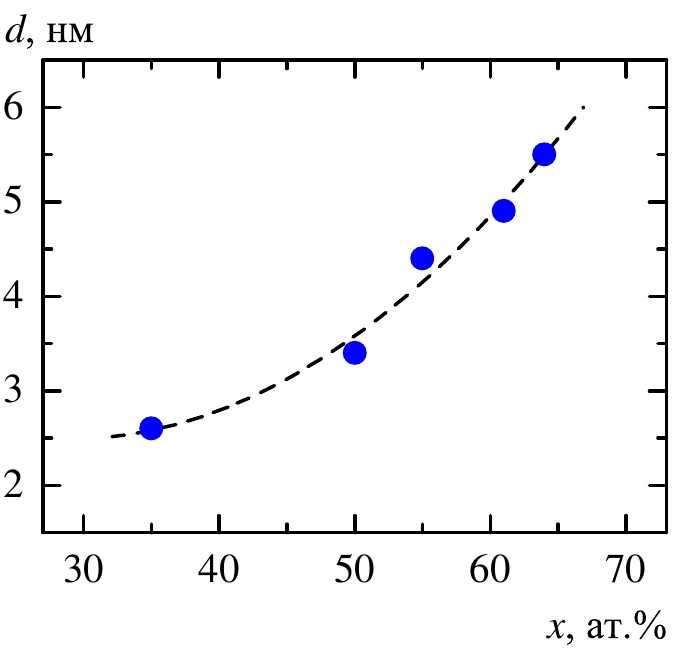}
\caption{\rm The average size of FM granules $d$ as a function of the FM phase content in the nanocomposite Ni$_x$(Al$_2$O$_3$)$_{100-x}$ according to X-ray diffraction data \cite{Stognei2016}.}
\end{figure}

The percolation threshold for all studied composites lies in the vicinity of metal phase concentrations $x\sim50$~at.\,\%. According to magnetic data, the transition of samples from superparamagnetic to ferromagnetic behavior occurs approximately in the same concentration range, or slightly lower \cite{Denisova2018, Stognei2014, Kalinin2007, Rylkov2017, Stognei2010, Tregubov2013, Stognei2016}.

In this work, the nanocomposite samples are studied by magnetic resonance in a wide range of frequencies ($f=7{-}37$~GHz) and temperatures ($T=4.2{-}360$~K) using a laboratory transmission-type spectrometer based on rectangular and tunable cylindrical resonators \cite{Drov2022-1}. In the used experimental geometry, the external static magnetic field $\mathbf{H}$ (up to 17~kOe) lies in the film plane. At the same time, it is possible to orient the high-frequency magnetic field $\mathbf{h}$ either perpendicular ($\mathbf{h}\perp\mathbf{H}$) or parallel ($\mathbf{h}\parallel\mathbf{H}$) to the static field H (``transverse'' and ``longitudinal'' geometry of resonance excitation).

\vskip 5mm
\noindent
\textbf{3. Experimental results}
\vskip 2mm

Figures~2,\,3 show experimental magnetic resonance spectra obtained for different nanocomposites M$_x$D$_{100-x}$ at room temperature. In the conventional transverse resonance excitation geometry ($\mathbf{h}\perp\mathbf{H}$), an intense FMR absorption peak is observed for all structures. As it was shown earlier \cite{DrovJMMM, Drov2022-1, Drov2023}, the frequency-field diagrams for this peak, as well as the dependencies of its position on the orientation of the field with respect to the film plane, are well described by the usual Kittel formulas, taking into account the demagnetizing factor $4\pi M$.

\begin{figure}
\centering
\includegraphics[width=0.85\columnwidth]{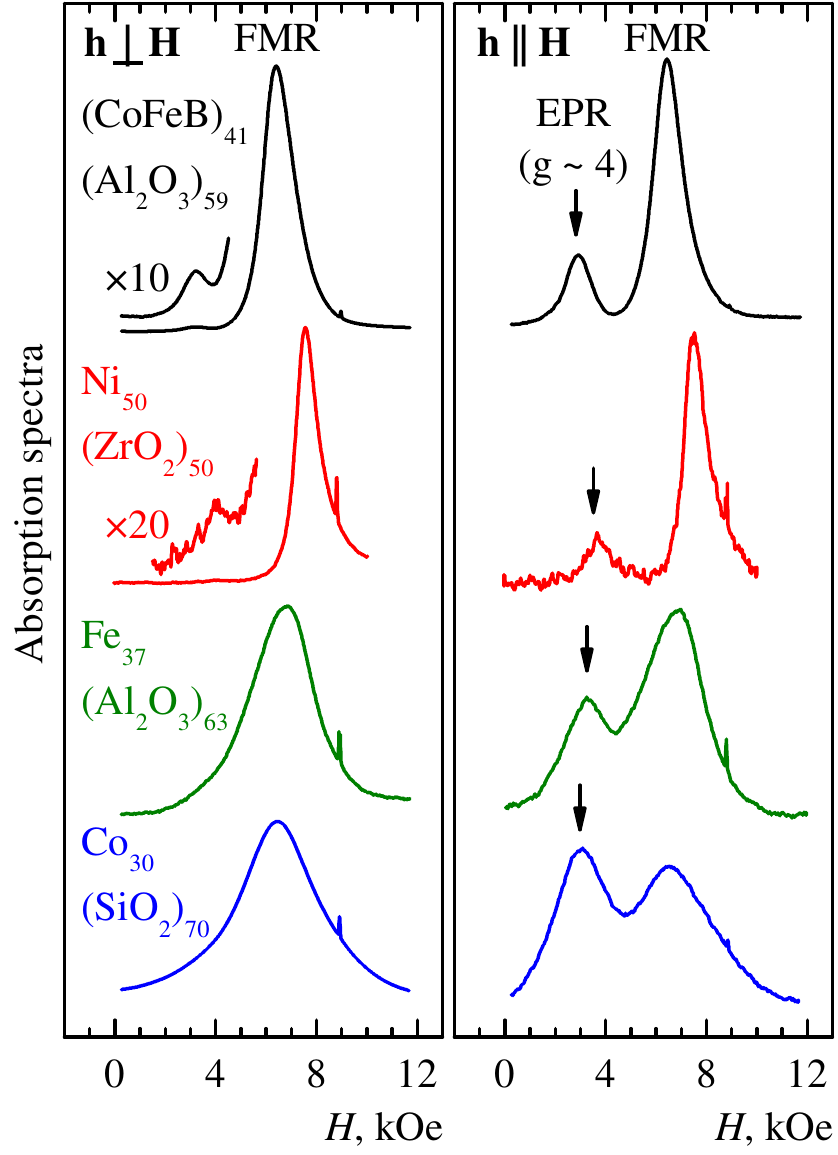}
\caption{\rm Magnetic resonance spectra for nanocomposite films with various compositions M$_x$D$_{100-x}$ in transverse ($\mathbf{h}\perp\mathbf{H}$) and longitudinal ($\mathbf{h}\parallel\mathbf{H}$) excitation geometry. The spectra are obtained at frequency $f\approx25$~GHz at room temperature.}
\end{figure}

\begin{figure}
\centering
\includegraphics[width=0.83\columnwidth]{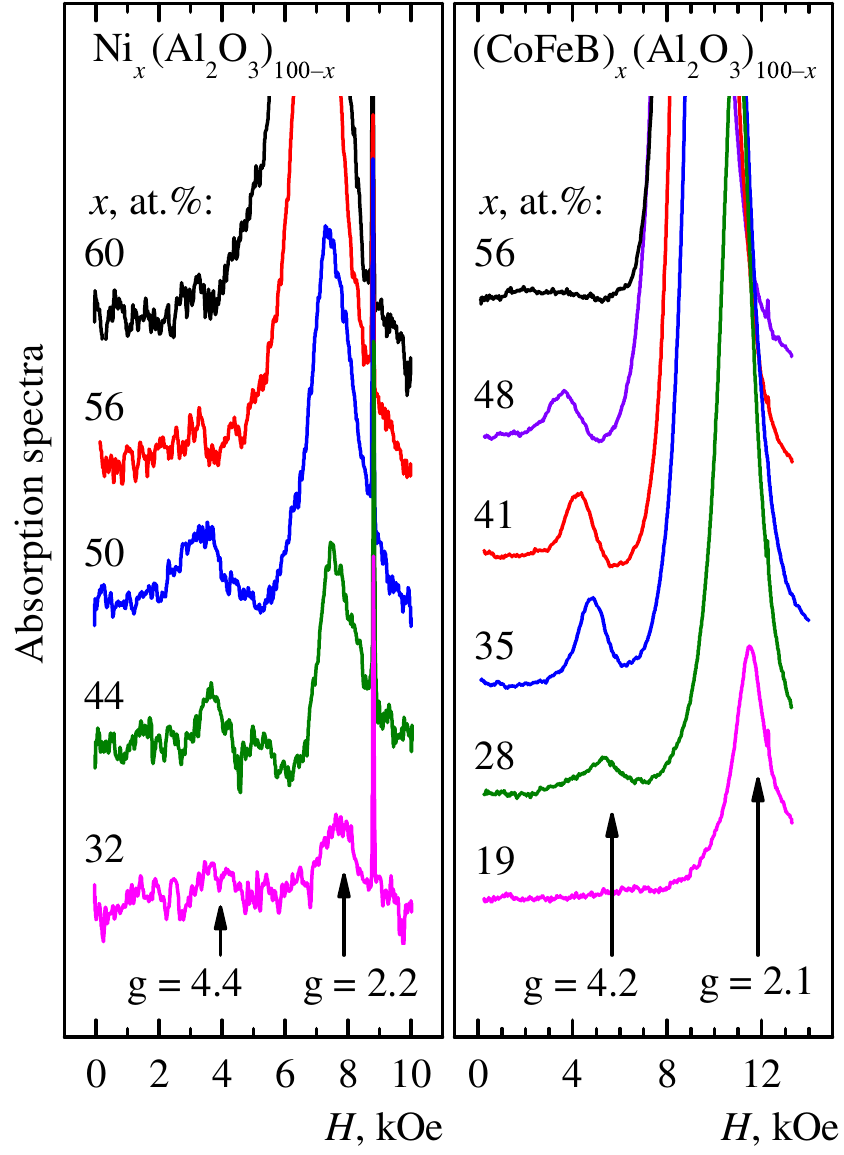}
\caption{\rm Experimental spectra for nanocomposite films Ni$_x$(Al$_2$O$_3$)$_{100-x}$ at frequency $f\approx24.6$~GHz and (CoFeB)$_x$(Al$_2$O$_3$)$_{100-x}$ ($f\approx34.4$~GHz) obtained at room temperature in the longitudinal geometry of resonance excitation ($\mathbf{h}\parallel\mathbf{H}$).}
\end{figure}

When the resonance excitation geometry is changed to a longitudinal one ($\mathbf{h}\parallel\mathbf{H}$), the intensity of the FMR peak decreases significantly. At the same time, a second absorption peak appears in weaker fields. As it was demonstrated in \cite{Drov2022-1, Drov2023}, the frequency-field and orientational dependencies for this peak are well described by a simple EPR formula
\begin{equation}
f = \gamma H_\mathrm{eff},
\end{equation}
where the gyromagnetic ratio $\gamma$ corresponds to the effective g-factor $g=4.3\pm0.1$, and the effective field $H_\mathrm{eff}$, in addition to the external field, includes dipolar fields produced inside the film by an array of FM granules.

Note that the $g\approx4$ EPR peak is manifested for systems of different compositions, including nanocomposites based on pure Co and Ni (Figs.~2,\,3), hence this peak cannot be associated with the well known EPR of isolated Fe$^{3+}$ ions ($g\approx4.3$) \cite{Koksharov2001, Jitianu2002, Edelman2012, Castner1960, Wickman1965, Kliava1988}. At the same time, it can be noted that the effective g-factors observed for the FMR and EPR peaks differ by about 2 times. For the FMR line, depending on the composition of the films, the g-factor varies in the range $g\approx2.1{-}2.2$, which is typical for metallic Fe, Co, Ni and their alloys \cite{Schoen2017}. For the EPR line, this value proves to be about 2 times larger $g\approx4.2{-}4.4$.

As an example, Fig.~4 shows frequency-field diagrams for two resonance peaks in Ni-based nanocomposites. In the high frequency range, the dependences $f(H)$ are described by linear functions, the slope of which differs by a factor 2.

Note that the narrowest resonance peaks are observed for (CoFeB)$_x$(Al$_2$O$_3$)$_{100-x}$ and Ni$_x$D$_{100-x}$ films. For these structures, the $g\approx4$ EPR peak can be resolved in both longitudinal and transverse resonance excitation geometry (Fig.~2), and its amplitude proves to be about the same in both geometries \cite{Drov2022-1, Drov2023}.

Another unusual feature of the $g\approx4$ EPR peak is the anomalous dependence of its intensity on the content of the FM phase in the nanocomposite (Fig.~3) and on temperature (Figs.~5--7). Note that in this regard, the main FMR peak behaves quite naturally. Its intensity rises monotonically with an increase of the FM phase concentration and with a decrease of temperature. A~different behavior is observed for the EPR peak. The spectra measured at room temperature demonstrate a non-monotonic dependence of the $g\approx4$ EPR peak amplitude on the content of the FM phase $x$ in the nanocomposite (Fig.~3). With increasing~$x$, the peak intensity first rises, however, when passing beyond the percolation threshold $x\gtrsim50$~at.\,\%, the peak decreases and disappears.

As the content of the FM phase in the nanocomposite rises, the character of the temperature dependence of the EPR peak intensity also changes (Figs.~5--7). At~low FM phase concentrations $x\lesssim25$~at.\,\%, the peak intensity $I(T)$ increases monotonously with temperature decrease, according to the conventional Curie law $I(T) \propto 1/T$. However, as $x$ rises, the dependence $I(T)$ becomes non-monotonic with a maximum that shifts to higher temperatures at growing $x$. Finally, in the limit of high FM phase concentrations, only an approximately linear growth $I(T) \propto T$ is observed (Fig.~6) \cite{Drov2022-2}.

\begin{figure}
\centering
\includegraphics[width=0.901\columnwidth]{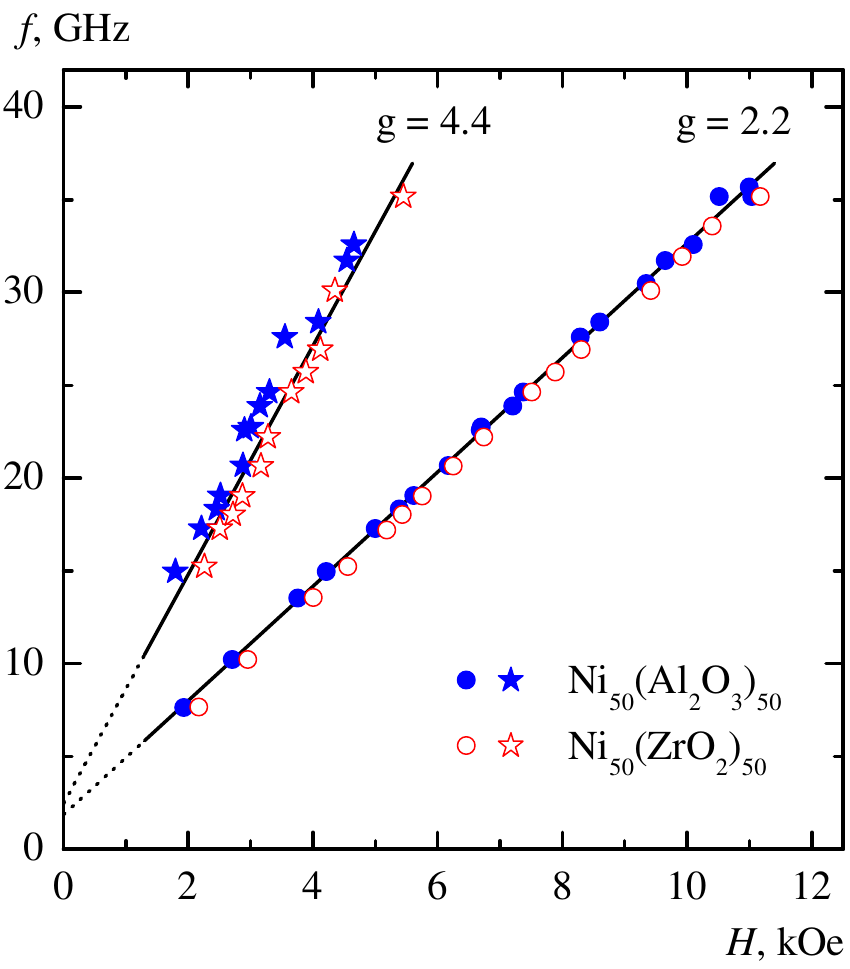}
\caption{\rm Frequency-field dependencies for two resonance peaks in films Ni$_{50}$(Al$_2$O$_3$)$_{50}$ and Ni$_{50}$(ZrO$_2$)$_{50}$ at room temperature. Symbols are experimental data, the slopes of the straight lines correspond to the effective g-factors $g=2.2$ and $g=4.4$.}
\end{figure}

\vskip 5mm
\noindent
\textbf{4. The ``giant spin'' model}
\vskip 2mm

The appearance of a peak with a double effective g-factor in the resonance spectra can be explained within the framework of the ``giant spin'' model by excitation of ``double quantum'' transitions inside FM nanogranules with a change of the spin projection $\Delta m = \pm2$ \cite{Noginova2008, Noginova2016, Fittipaldi2016, Fittipaldi2017, Domracheva2017}. Within the framework of this approach, it is possible to explain the better manifestation of the $g\approx4$ peak in the longitudinal geometry of the resonance excitation \cite{Fittipaldi2016, Fittipaldi2017, Domracheva2017}, as well as the anomalous temperature dependence of its intensity \cite{Noginova2008, Fittipaldi2016, Fittipaldi2017}.

\begin{figure}
\centering
\includegraphics[width=0.901\columnwidth]{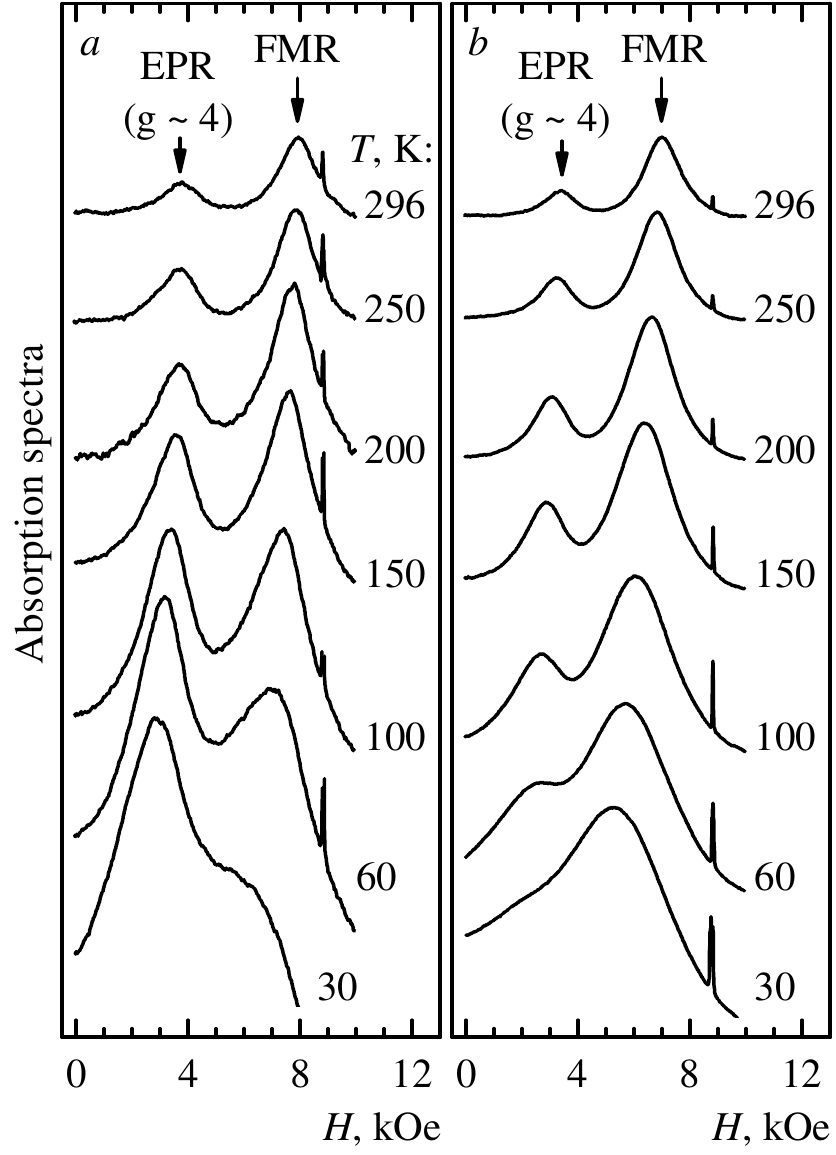}
\caption{\rm Spectra of the films (CoFeB)$_x$(Al$_2$O$_3$)$_{100-x}$ with $x\approx25$~at.\,\% (\textit{a}) and $35$~at.\,\% (\textit{b}) at different temperatures $T=4.2{-}296$~K, obtained at frequency $f\approx25$~GHz in longitudinal resonance excitation geometry ($\mathbf{h}\parallel\mathbf{H}$).}
\end{figure}

In the giant spin model, the FM nanogranule is treated as a PM center with a very large spin $S \sim 10^2{-}10^4$. In the external field, Zeeman splitting of energy levels occurs, according to the spin projection on the field direction $m = -S\dotsc{+}S$. The transitions with a change of the spin projection $\Delta m = \pm1$ induced by an alternating field in the classical limit correspond to the excitation of the usual FMR mode. Formally ``forbidden'' (``double quantum'') transitions with $\Delta m = \pm2$ become ``allowed'' taking into account additional perturbations in the system, for example, in the presence of magnetic anisotropy of granules, or dipole-dipole interactions between them \cite{Abragam1972, Altshuler1972}.

Let us consider the simplest case of a weak uniaxial anisotropy in the granules with effective field $H_A$. Such anisotropy can arise in the granules, for example, when their shape deviates from spherical. In this case, the quantum mechanical probability $f_{m\pm1}$ of transitions between the levels $m\pm1$ in the framework of perturbation theory is estimated by the expression \cite{Marti, Clerjaud}
\begin{equation}
\begin{split}
&f_{m\pm1} \propto \left( \frac{H_A}{H} \right)^2 \frac{(U_{m}^{m-1} U_{m+1}^{m})^2}{S^2},\\
&\text{where ~} U_m^{m-1} = \sqrt{S(S+1)-m(m-1)}.
\end{split}
\end{equation}

Note that for an arbitrary orientation of the anisotropy axis relative to the direction of the field, the probabilities of such transitions excited by transverse and longitudinal microwave field turn out to be comparable \cite{Clerjaud}, which result is consistent with the experimentally observed behavior for the $g\approx4$ EPR peak.

It is important, that according to equation (2), the probability of transitions with $\Delta m = \pm2$ near the ground state of the granule $m=-S$ vanishes. On the contrary, the maximum probabilities of these transitions are realized at low m values $|m| \ll S$. However, the corresponding energy levels lie above the ground state $m=-S$ by an amount of $\mu H$, where $\mu$ is the granule magnetic moment. Therefore, at low temperatures ($k_BT \ll \mu H$, where $k_B$ is Boltzmann constant), when the granules pass into the ground state, the intensity of the double quantum line $I(T)$ decreases. In the high temperature limit ($k_BT \gg \mu H$), the populations of the energy levels in the granules tend to equalize leading to the conventional Curie law $I(T) \propto 1/T$. Thus, the maximum intensity of the $g\approx4$ line is expected at $T \sim \mu H/k_B$. With an increase of the FM phase concentration in the nanocomposite, the magnetic moments of the granules increase leading to the shift of the maximum $I(T)$ to higher temperatures. Above the percolation threshold, the granules begin to form macroscopic clusters, $\mu$ increases sharply, and the double quantum line disappears completely.

\begin{figure}
\centering
\includegraphics[width=0.829\columnwidth]{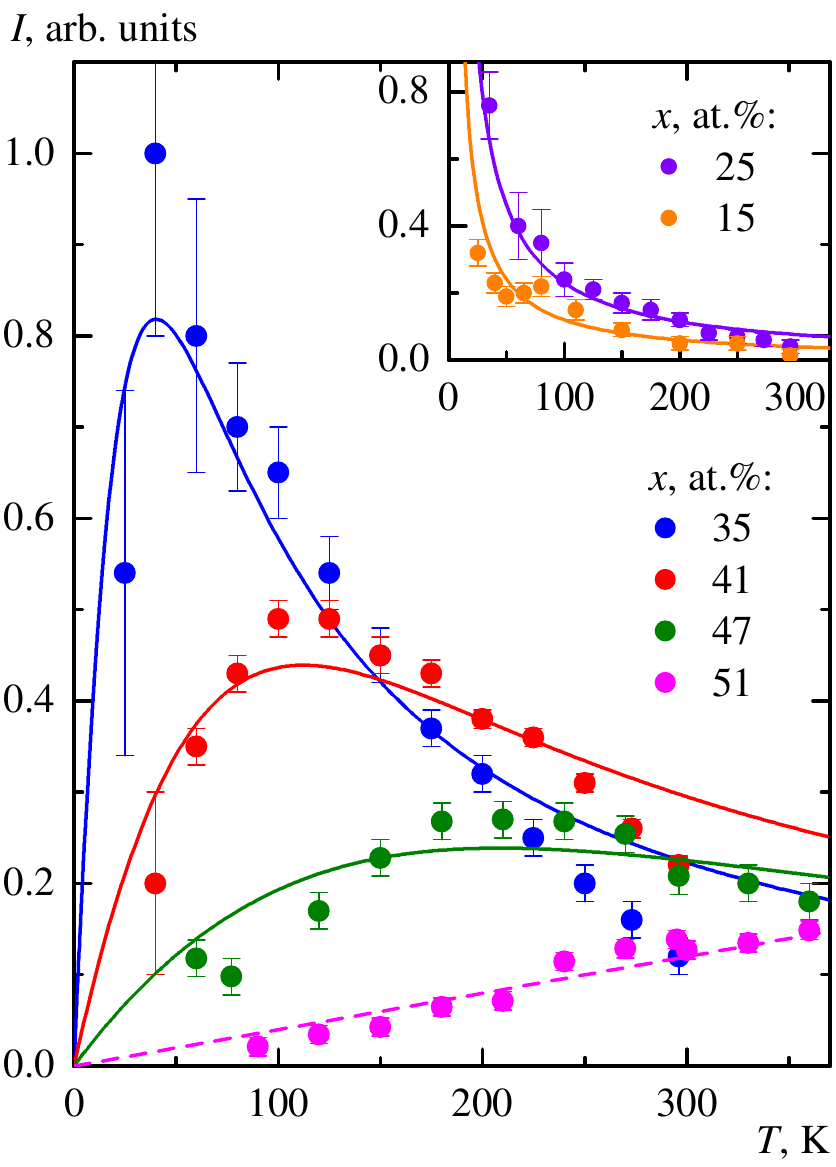}
\caption{\rm Temperature dependencies of the integral intensity $I(T)$ for the $g\approx4$ EPR line in nanocomposite films (CoFeB)$_x$(Al$_2$O$_3$)$_{100-x}$. Symbols are experimental data at frequency $f\approx25$~GHz, solid lines are calculations in the “giant spin” model, the dashed line is linear dependence. Lines in the inset are the Curie law $I(T)\propto1/T$.}
\end{figure}

To estimate the intensity of the line quantitatively, we take into account the difference in the populations of the levels $m\pm1$ at finite temperature $T$:
$$
\Delta \rho_{m\pm1}(T) = Z_S^{-1} \left( e^{-2(m-1)\varkappa} - e^{-2(m+1)\varkappa} \right),
$$
where $Z_S$ is the statistical sum given by the well-known expression \cite{Smart}
$$
Z_S = \frac{\mathrm{sh}[(2S+1)\varkappa]}{\mathrm{sh}(\varkappa)}, \quad \varkappa = \frac{\mu_B H_\mathrm{eff}}{k_B T}
$$
($\mu_B$ is Bohr magneton). The effective field $H_\mathrm{eff}$ in the above equations, besides the external field, includes magnetic dipolar fields created inside the film by an array of FM granules. According to equation (1), this field is defined by the simple ratio $H_\mathrm{eff}=f/\gamma$. Taking into account the probabilities of double quantum transitions (2), the integral intensity of the $g\approx4$ EPR line is proportional to
\begin{equation}
I(T) \propto \sum_{m=1-S}^{S-1} f_{m\pm1} \Delta \rho_{m\pm1}(T).
\end{equation}

Figures~6,\,7 show the experimental dependencies $I(T)$ for the $g\approx4$ EPR peak in films with different compositions comparing with the results of numerical calculations in the described model. The fitting parameters are the vertical scale of the function (3) and the spin of the particle $S$ (the corresponding magnetic moment $\mu=2S\mu_B$), which determines the position of the maximum $I(T)$.

\begin{figure}
\centering
\includegraphics[width=0.871\columnwidth]{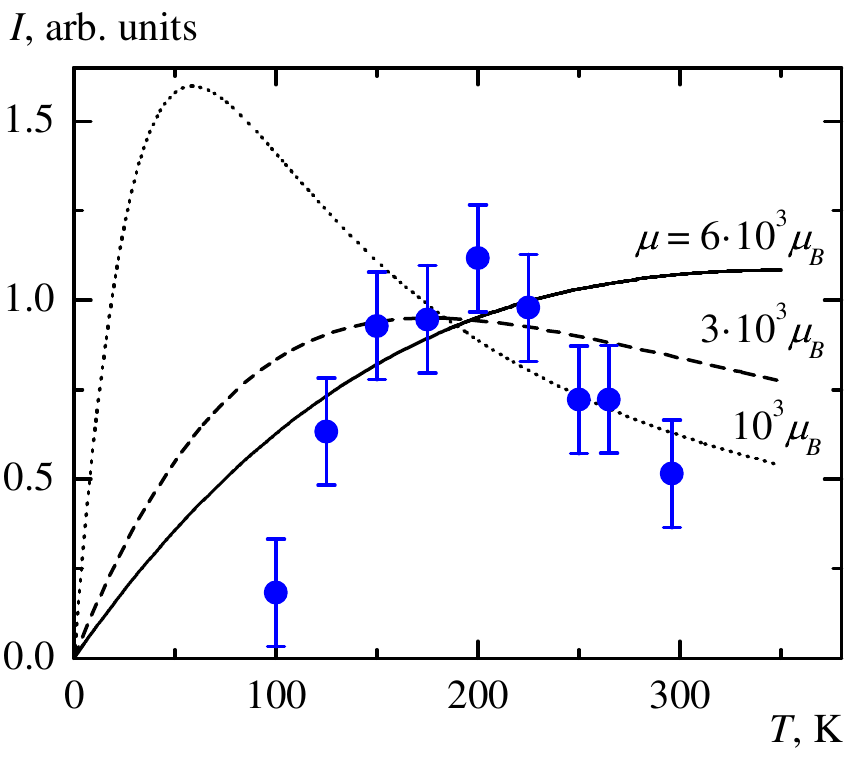}
\caption{\rm Temperature dependencies of the integral intensity $I(T)$ for the $g\approx4$ EPR line in nanocomposite film Ni$_{50}$(Al$_2$O$_3$)$_{50}$. Symbols are experimental data at frequency $f\approx18$~GHz, lines are calculations in the ``giant spin'' model with different values of granule magnetic moment $\mu$.}
\end{figure}

\begin{figure}
\centering
\includegraphics[width=0.84\columnwidth]{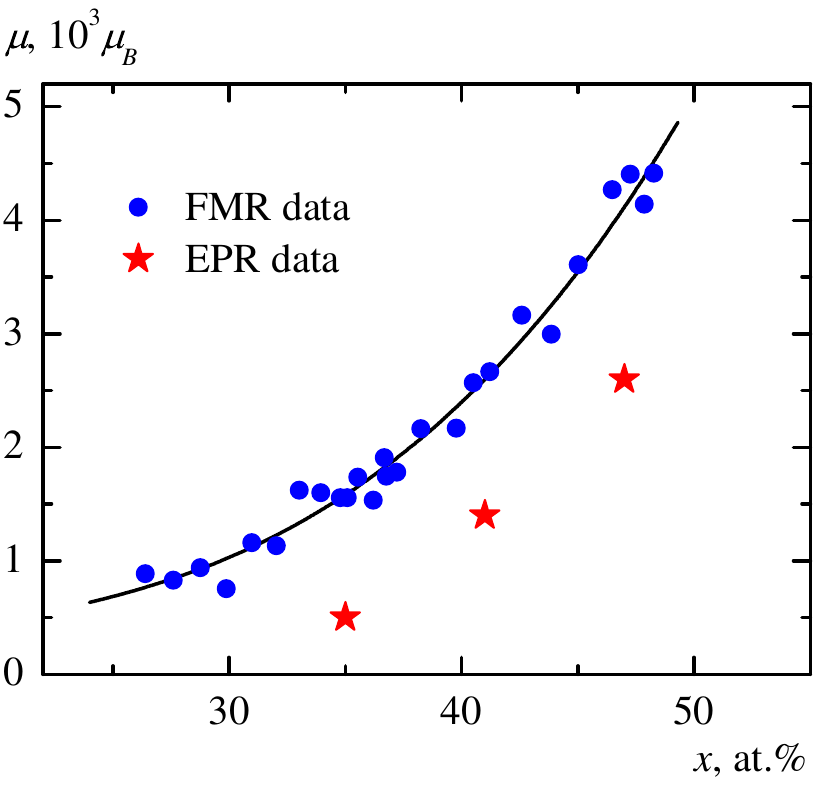}
\caption{\rm The granule magnetic moment $\mu$ as a function of the FM phase concentration in nanocomposite (CoFeB)$_x$(Al$_2$O$_3$)$_{100-x}$, obtained from the approximation of the curves $4\pi M(H)$ (according to FMR data) and from temperature dependences of the EPR peak intensity.}
\end{figure}

On a qualitative level, the experimental data are consistent with the theory. At the same time, systematic quantitative discrepancies can also be noticed. Comparing with the calculation, the experimental dependences $I(T)$ demonstrate more pronounced maxima. This discrepancy is more evident for the Ni-based nanocomposite (Fig.~7). As shown in Fig.~7, the observed behavior can be formally attributed to a decrease of the granule magnetic moment $\mu$ with increasing temperature. Indeed, taking into account the finite Curie temperature $T_C$ of the granules, under the condition $T\lesssim T_C$, the effect of reducing $\mu(T)$ is quite expected. In the case of Ni granules, this effect should be more pronounced due to the lower $T_C$ compared to CoFeB. It should also be noted that with temperature decrease, magnetic interactions between nanogranules can play an important role and lead to the formation of larger magnetically ordered clusters, causing an additional increase in the effective value $\mu$ \cite{Rylkov2023}. However, the considered simplest model completely neglects the effects of intergranular interactions and does not take into account the presence of excited states of nanoparticles with a reduced value of the total spin at $T\lesssim T_C$, and therefore is only a qualitative approximation of the real situation.

Figure~8 shows the results of estimation of the granule magnetic moment $\mu$ from the dependencies $I(T)$ for the $g\approx4$ EPR peak in films (CoFeB)$_x$(Al$_2$O$_3$)$_{100-x}$. For comparison, Fig.~8 also demonstrates the values $\mu$ obtained from approximation of experimental curves $4\pi M(H)$ by the Langevin function (the dependencies $4\pi M(H)$ were extracted from FMR data at room temperature, as described in the work \cite{Drov2023}). As can be seen, the magnetic moments of the granules, defined in two ways, prove to have the same order. However, EPR data systematically show lower $\mu$ values compared to FMR data. It can be assumed that this difference is due to the dispersion of the FM granules in size. In this case, the FMR peak is mainly determined by resonance in large granules, while the $g\approx4$ EPR peak is more efficiently excited in small magnetic particles.

\vskip 5mm
\noindent
\textbf{5. Conclusion}
\vskip 2mm

Films of metal-insulator nanogranular composites M$_x$D$_{100-x}$ with various compositions (M~= Fe, Co, Ni, CoFeB; D = Al$_2$O$_3$, SiO$_2$, ZrO$_2$) and contents of the metallic FM phase $x\approx15{-}60$~at.\,\% were studied by electron magnetic resonance. The experimental spectra, in addition to the conventional FMR signal, contain an additional absorption peak with a double effective g-factor $g\approx4$ demonstrating a number of unusual features. The appearance of such a peak in the spectra and its properties can be explained within the framework of the quantum mechanical ``giant spin'' model, considering FM nanogranules as PM centers with very large spin $S\sim 10^2{-}10^4$. The observed $g\approx4$ EPR line is associated with the excitation of double quantum transitions in these PM centers with a change of the spin projection $\Delta m = \pm2$. The proposed approach makes it possible to explain qualitatively the non-monotonic temperature dependence of the peak intensity, the non-standard conditions for its excitation by longitudinal microwave magnetic field, as well as the disappearance of this peak at contents of the FM phase above the percolation threshold of the nanocomposite.

Thus, the observed special features of the magnetic resonance in nanogranular composites emphasize the ``dualism'' of classical and quantum properties of ferromagnetic nanoparticles. The behavior of the main FMR line is well described within the framework of the classical concept as the excitation of magnetization vector precession in an ensemble of magnetic nanogranules. At the same time, the properties of the additional peak are explained in quantum mechanical terms by the excitation of EPR transitions between the spin states of individual nanogranules.

\vskip 5mm
\noindent
\textbf{Funding:} The work was carried out within the framework of a state assignment and was financially supported by the Russian Science Foundation (project no. 22--19--00171).

\end{document}